\documentclass[aps,twocolumn,prl]{revtex4}

\usepackage{amsmath}
\usepackage{bm}
\usepackage{dcolumn}
\usepackage{psfig}
\usepackage{graphicx}

\begin{document}

\title{Determination of the Charged Pion Form Factor at $Q^2$=1.60 and 2.45 (GeV/c)$^2$}

\author{
T.~Horn,$^{1}$ 
K.~Aniol,$^{3}$
J.~Arrington,$^{2}$ 
B.~Barrett,$^{4}$ 
E.J.~Beise,$^{1}$
H.P.~Blok,$^{5, 6}$
W.~Boeglin,$^{7}$
E.J.~Brash,$^{8}$ 
H.~Breuer,$^{1}$ 
C.C.~Chang,$^{1}$ 
M.E.~Christy,$^{9}$ 
R.~Ent,$^{10}$ 
D.~Gaskell,$^{10}$ 
E.~Gibson,$^{11}$ 
R.J.~Holt,$^{2}$ 
G.M.~Huber,$^{8}$ 
S.~Jin,$^{12}$ 
M.K.~Jones,$^{10}$ 
C.E.~Keppel,$^{9, 10}$
W.~Kim,$^{12}$
P.M.~King,$^{1}$ 
V.~Kovaltchouk,$^{8}$ 
J.~Liu,$^{1}$ 
G.J.~Lolos,$^{8}$ 
D.J.~Mack,$^{10}$
D.J.~Margaziotis,$^{3}$
P.~Markowitz,$^{7}$ 
A.~Matsumura,$^{13}$ 
D.~Meekins,$^{10}$
T.~Miyoshi,$^{13}$ 
H.~Mkrtchyan,$^{14}$ 
I.~Niculescu,$^{15}$ 
Y.~Okayasu,$^{13}$
L.~Pentchev,$^{16}$ 
C.~Perdrisat,$^{16}$ 
D.~Potterveld,$^{2}$ 
V.~Punjabi,$^{17}$
P.~Reimer,$^{2}$ 
J.~Reinhold,$^{7}$
J.~Roche,$^{10}$ 
P.G.~Roos,$^{1}$
A.~Sarty,$^{4}$ 
G.R.~Smith,$^{10}$ 
V.~Tadevosyan,$^{14}$ 
L.G.~Tang,$^{9, 10}$ 
V.~Tvaskis,$^{5, 6}$ 
S.~Vidakovic,$^{8}$
J.~Volmer,$^{18}$ 
W.~Vulcan,$^{10}$
G.~Warren,$^{10}$
S.A.~Wood,$^{10}$ 
C.~Xu,$^{8}$ and
X.~Zheng$^{2}$\\
(The Jefferson Lab F$_{\pi}$-2 Collaboration)
}
\affiliation{$^{1}$ Department of Physics, University of Maryland, College Park, Maryland 20742}
\affiliation{$^{2}$ Argonne National Laboratory, Argonne, Illinois 60439}
\affiliation{$^{3}$ California State University Los Angeles, Los Angeles, California 90032}
\affiliation{$^{4}$ Saint Mary's University, Halifax, Nova Scotia, Canada B3H 3C3}
\affiliation{$^{5}$ Vrije Universiteit, NL-1081 HV Amsterdam, The Netherlands}
\affiliation{$^{6}$ NIKHEF, Postbus 41882, NL-1009 DB Amsterdam, The Netherlands}
\affiliation{$^{7}$ Florida International University, University Park, Florida 33199}
\affiliation{$^{8}$ University of Regina, Regina, SK, Canada S4S 0A2}
\affiliation{$^{9}$ Hampton University, Hampton, Virginia 23668}
\affiliation{$^{10}$ Thomas Jefferson National Accelerator Facility, Newport News, Virginia 23606}
\affiliation{$^{11}$ California State University, Sacramento, California 95819}
\affiliation{$^{12}$ Kyungook National University, Taegu, Korea}
\affiliation{$^{13}$ Tohuku University, Sendai, Japan}
\affiliation{$^{14}$ Yerevan Physics Institute, Yerevan, Armenia}
\affiliation{$^{15}$ James Madison University, Harrisonburg, Virginia 22807}
\affiliation{$^{16}$ College of William and Mary, Williamsburg, Virginia 23187}
\affiliation{$^{17}$ Norfolk State University, Norfolk, Virginia 23504}
\affiliation{$^{18}$ DESY, Hamburg, Germany}
\date{\today}

\begin{abstract}
The $^{1}$H($e,e^\prime \pi^+$)n cross section was measured at four-momentum transfers 
of $Q^2$=1.60 and 2.45 GeV$^2$ at an invariant mass of the photon nucleon system of $W$=2.22 GeV. The charged pion form factor ($F_{\pi}$) was extracted from the data by comparing the separated longitudinal pion electroproduction cross section to a Regge model prediction in which $F_{\pi}$ is a free parameter. The results indicate that the pion form factor deviates from the charge-radius constrained monopole form at these values of $Q^2$ by one sigma, but is still far from its perturbative Quantum Chromo-Dynamics prediction.
\end{abstract}

\maketitle

A fundamental challenge in nuclear physics is the description of hadrons in terms of the constituents of the underlying theory of strong interactions, quarks and gluons. Properties such as the total charge and magnetic moments are well described in a constituent quark framework, which effectively takes into account quark-gluon interactions. However, charge and current distributions, which are more sensitive to the underlying dynamic processes, are not well described. 

Hadronic form factors provide important information about hadronic structure. The coupling of a virtual photon to structureless particles is completely determined by their charge and magnetic moments. However, for composite particles one must account for the internal structure, which is accomplished by momentum transfer dependent functions. Examples of these functions are the electromagnetic form factors, which describe the distribution of charge and current.  

One of the simplest hadronic systems available for study is the pion, whose valence structure is a bound state of a quark and an antiquark. The electromagnetic structure of a spinless particle such as the pion is parameterized by a single form factor. The pion charge form factor, $F_{\pi}$, can be calculated in perturbative Quantum Chromo-Dynamics (pQCD) in the limit of very large values of four-momentum transfer squared, $Q^2$ \cite{Farr79}:
\begin{equation}                                                               
    F_{\pi}\left(Q^2\right) = 8 \pi \ \frac{\alpha_s \ f^2_{\pi}}{Q^2}  \ \ \ \ (Q^2 \rightarrow \infty), 
\end{equation}  
where $\alpha_s$ is the strong coupling constant. The normalization is fixed by the pion decay constant, $f_{\pi}$, which is determined from the weak decay of the pion ($\pi \rightarrow \mu + \nu_{\mu}$). At low values of $Q^2$, Vector Meson Dominance models provide a reasonable description of $F_{\pi}$. Due to the pion's simple $\bar{q}q$ valence structure, the transition from ``soft'' (nonperturbative) to ``hard'' (perturbative) physics is expected to occur at significantly lower values of $Q^2$ for $F_{\pi}$ than for the nucleon form factors \cite{Isg84}. 

The form factor of the pion is well determined up to $Q^2$=0.28 GeV$^2$ by elastic $\pi-e$ scattering experiments \cite{Ame84}, from which the charge radius has been extracted. Extending the measurement of $F_{\pi}$ to larger values of $Q^2$ requires the use of pion electroproduction from a nucleon target. The pion exchange ($t$-pole) process, in which a virtual photon couples to a virtual pion inside the nucleon, dominates the longitudinal pion electroproduction cross section, $\sigma_L$, at small values of the Mandelstam variable $t$. There $\sigma_L$ exhibits a characteristic $t$-dependence and is proportional to $F^2_{\pi}$ .

Experimental values of $F_{\pi}$ have previously been determined at CEA and Cornell \cite{Beb76, Beb78}, DESY \cite{Bra77, Ack78}, and recently at Jefferson Lab \cite{Vol01}. Most of the high $Q^2$ data have come from experiments at Cornell covering a range of values in $Q^2$ between 0.28 and 9.77 GeV$^2$. In these experiments $F_{\pi}$ was extracted from the longitudinal cross sections, which were isolated by subtracting a model of the transverse contribution from the unseparated cross sections. Pion electroproduction data were also obtained at DESY for a value of $Q^2$ of 0.7 GeV$^2$, at an invariant mass of the photon nucleon system of $W$=2.19 GeV, and longitudinal and transverse cross sections were extracted using the Rosenbluth separation method. In 1997, Jefferson Lab provided the first high precision pion electroproduction data for $F_{\pi}$ for values of $Q^2$ between $Q^2$=0.6 and 1.6 GeV$^2$ at $W$=1.95 GeV \cite{Vol01}. For an updated analysis of these data see reference \cite{Tad06}. These data give a precise determination of $\sigma_L$ with a significant improvement in the experimental uncertainty. The results presented here extend the $Q^2$ range to 2.45 GeV$^2$ and address questions of model dependence in the extraction of $F_{\pi}$. 

The experiment described here was carried out in Hall C at Thomas Jefferson National Accelerator Facility (Jefferson Lab). Pion electroproduction cross sections were measured from hydrogen and deuterium targets. The data were taken at two beam energies for each of the two values of $Q^2$ at $W$=2.22 GeV. Charged pions were detected in the High Momentum Spectrometer (HMS), while the scattered electrons were detected in the Short Orbit Spectrometer (SOS). 
\begin{figure}
\begin{center}
\includegraphics[width=3.0in]{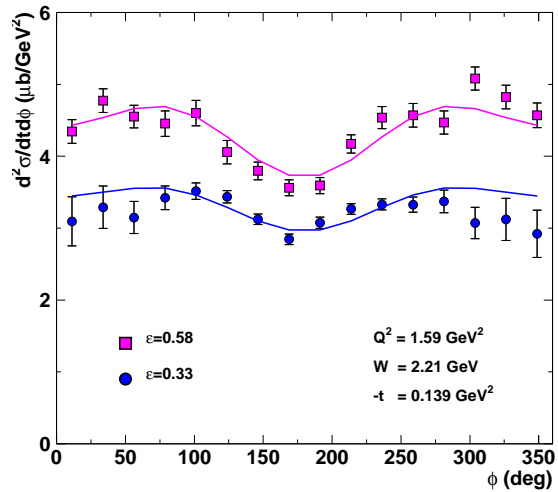}
\caption{\label{fig-unsep-xsec} \it Representative example of the measured cross sections, $\frac{d^2 \sigma}{dt d\phi}$ as a function of $\phi$ at $Q^2$=1.6 GeV$^2$ for two values of $\epsilon$. The curves shown represent the model cross section used in the Monte Carlo simulation.}
\end{center}
\end{figure}
Both spectrometers include two drift chambers for track reconstruction and scintillator arrays for triggering. A detailed description of the Jefferson Lab Hall C spectrometers can be found in reference \cite{Horn06}. 

In order to select electrons in the SOS, a gas \v{C}erenkov detector containing Freon-12 at atmospheric pressure was used in combination with a lead-glass calorimeter. Positively charged pions were identified in the HMS using an aerogel \v{C}erenkov detector with refractive index of 1.03 \cite{Asa05}. In the case of pion production at negative polarity, electrons were rejected using a gas \v{C}erenkov detector containing C$_4$F$_{10}$ at 0.47 atm. Any remaining contamination from real electron-proton coincidences was eliminated with a coincidence time cut of $\pm$1 ns. Background from alumininum target cell walls (2-4\% of the yield) and random coincidences ($\sim$1\%) were subtracted from the charge normalized yields. The exclusive neutron final state was selected with a cut on the reconstructed missing mass. The relevant electroproduction kinematic variables $Q^2, W$ and $t$ were reconstructed from the measured spectrometer quantities. Experimental yields were calculated after correcting for several inefficiencies, the dominant sources being particle tracking efficiency (3-4\%), pion absorption (4.8\%), and computer dead time (1-11\%). The net uncertainty in these corrections is dominated by the uncertainty in the absorption of the pions ($\sim$2\%).

The unpolarized pion electroproduction cross section can be written as the product of a virtual photon flux factor and a virtual photon cross section,
\begin{equation}
   \frac{d^5 \sigma}{d \Omega_e  dE_e^\prime  d \Omega_{\pi}} = J\left(t,\phi \rightarrow \Omega_{\pi}\right) \Gamma_v \frac{d^2 \sigma}{dt d \phi},
\end{equation}
where $J\left(t,\phi \rightarrow \Omega_{\pi}\right)$ is the Jacobian of the transformation from $dtd\phi$ to $d\Omega_{\pi}$, $\phi$ is the azimuthal angle between the scattering and the reaction plane, and $\Gamma_v$=$\frac{\alpha}{2 \pi^2} \frac{E^\prime_e}{E_e} \frac{1}{Q^2} \frac{1}{1-\epsilon} \frac{W^2-M^2}{2 M}$ is the virtual photon flux factor. The virtual photon cross section can be expressed in terms of contributions from transversely and longitudinally polarized photons,
\begin{eqnarray}
\label{eqn-unsep}
  2\pi \frac{d^2 \sigma}{dt d\phi} & = & \frac{d \sigma_T}{dt} + \epsilon  \frac{d \sigma_L}{dt} 
                                    +  \sqrt{2 \epsilon (1 + \epsilon)}  \frac{d \sigma_{LT}}{dt} cos \phi \\ \nonumber
                                   & + & \epsilon  \frac{d \sigma_{TT}}{dt} cos 2 \phi.
\end{eqnarray}
Here, $\epsilon=\left(1+2\frac{|{\bf q^2}|}{Q^2}\tan^2\frac{\theta}{2}\right)^{-1}$ is the virtual photon polarization, where ${\bf q^2}$ is the square of the three-momentum transferred to the nucleon and $\theta$ is the electron scattering angle. The individual components in equation~\ref{eqn-unsep} were determined from a simultaneous fit to the $\phi$ dependence of the measured cross sections, $\frac{d^2 \sigma}{dt d\phi}$, at two values of $\epsilon$. A representative example as a function of $\phi$ is shown in Figure~\ref{fig-unsep-xsec}.  

\begin{figure}
\begin{center}
\includegraphics[width=3.5in]{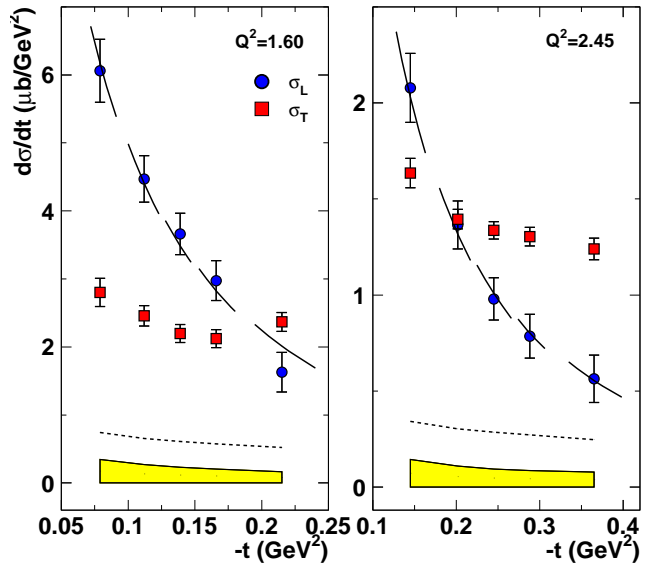}
\caption{\label{fig-sep-xsec} \it Separated cross sections, $\sigma_L$ and $\sigma_T$ at central values of $Q^2$=1.60 (2.45) GeV$^2$. Note that the average values of $W$ and $Q^2$ are different for each $-t$-bin. The error bar indicates statistical and uncorrelated systematic uncertainty in both $\epsilon$ and $-t$ combined in quadrature. The error band denote the correlated part of the systematic uncertainty by which all data points move collectively. The curves denote VGL Regge calculations for $\sigma_L$ (solid line) and $\sigma_T$ (dashed line) with values of $\Lambda^2_{\pi}$=0.513 (0.491) GeV$^2$ and $\Lambda^2_{\rho}$=1.1 GeV$^2$. The discontinuities in the $\sigma_L$ curve result from the different average values of $W$ and $Q^2$ for the various $t$-bins.}
\end{center}
\end{figure}

The separated cross sections are determined at fixed values of $W$, $Q^2$ and $t$, common for both high and low values of $\epsilon$. However, the acceptance covers a range in these quantities, thus the measured yields represent an average over that range. Note that each $t$-bin has a different average value of $Q^2$, $W$. In order to minimize errors resulting from averaging, the experimental cross sections were calculated by comparing the experimental yields to a Monte Carlo simulation of the experiment. To account for variations of the cross section across the acceptance the simulation uses a $^{1}$H(e,e$^\prime \pi^+$)n model based on pion electroproduction data. In addition, the Monte Carlo includes a detailed description of the spectrometers, multiple scattering, ionization energy loss, pion decay, and radiative processes. The separated cross sections, $\sigma_L$ and $\sigma_T$, are shown in Figure~\ref{fig-sep-xsec}.
      
The uncertainty in the separated cross sections has both statistical and systematic sources. The statistical uncertainty in $\sigma_T+\epsilon \sigma_L$ ranges between 1 and 2\%. Systematic uncertainties that are uncorrelated between high and low $\epsilon$ points are amplified by a factor of $1/\Delta \epsilon$ in the L-T separation. Correlated systematic uncertainties propagate directly into the separated cross sections. Uncertainties in the scattering kinematics and beam energy were parameterized using data from the over-constrained $^{1}$H($e,e^\prime p$) reaction. Beam energy and spectrometer momenta were determined to 0.1\% while the spectrometer angles were determined to $\approx$0.5 mrad. The spectrometer acceptance was verified to better than 2\% by comparing $e-p$ elastic scattering data to a global parameterization \cite{Arr04}. The uncorrelated systematic uncertainty is dominated by acceptance (0.6-1.1\%) resulting in a total uncorrelated uncertainty of 0.9 to 1.2\%. The correlated systematic uncertainty is mainly due to radiative corrections (2\%), pion absorption (2\%), and pion decay (1\%) resulting in a total correlated uncertainty of 3.5\%. A third category of systematic uncertainties consists of uncertainties that differ in size between $\epsilon$ points, but may influence the $t$-dependence at a fixed value of $\epsilon$ in a correlated way. The ``t-correlated'' uncertainty is dominated by acceptance (0.6\%), kinematics (0.8-1.1\%) and model dependence (1.1-1.3\%) resulting in a partially correlated uncertainty of 1.8 and 1.9\%.

In order to determine $F_{\pi}$, the experimental results for $\sigma_L$ are compared to a Regge model calculation by Vanderhaeghen, Guidal and Laget (VGL) \cite{Van98}. In this approach, pion electroproduction is described as the exchange of Regge trajectories for $\pi$- and $\rho$-like particles. Since most model parameters are fixed by pion photoproduction data, $F_{\pi}$ and the $\pi \rho \gamma$ transition form factor are the only free parameters. Both form factors are parameterized by a monopole form, $[1+Q^2/\Lambda^2_i]^{-1}$, but the cutoff parameter, $\Lambda^2_{\rho}$, is not as well constrained as the pion cutoff parameter, $\Lambda^2_{\pi}$. Varying $\Lambda^2_{\rho}$ between 0.6 and 2.1 GeV$^2$ changes $\sigma_T$ by 13\% (30\%) at $Q^2$ of 1.60 (2.45), but has little influence on $\sigma_L$. Thus, $F_{\pi}$ can be determined in a one parameter fit from a comparison of the longitudinal experimental cross section to the one predicted from the Regge model. 

A comparison of our data to the VGL prediction is shown in Figure~\ref{fig-sep-xsec}. The $t$-dependence of the longitudinal cross section is well described at both central values of $Q^2$. However, the transverse cross section is underpredicted systematically. The value of $F_{\pi}$ was determined from a least squares fit of the Regge model prediction to the data, and the resulting values are shown in Table~\ref{table_fpi}. 
  \begin{table}
 \begin{center}  
  \begin{tabular}{||c|c||}
  \hline  
  $Q^2$ (GeV$^2$)  & $F_{\pi}$  \\
 \hline
 \hline 
 1.60 & 0.243$\pm$0.012  \\
 2.45 & 0.167$\pm$0.010   \\  
  \hline  
  \end{tabular}
  \end{center}
 \caption{\label{table_fpi} \it Extracted values for $F_{\pi}$ at a value of $W$=2.22 GeV.  The error on $F_{\pi}$ combines statistical and experimental systematic uncertainties in quadrature.}
 \end{table}

The extraction of $F_{\pi}$ from $\sigma_L$ relies on the dominance of the pion exchange term. To test the pole dominance the longitudinal $\pi^-/\pi^+$ ratios in $^{2}$H were examined. Since the pole term is purely isovector this ratio is expected to be close to unity and a significant deviation from unity would indicate the presence of an isoscalar background. The preliminary analysis of the longitudinal $\pi^-$/$\pi^+$ ratios is consistent with unity.
\begin{figure}
\begin{center}
\includegraphics[width=3.5in]{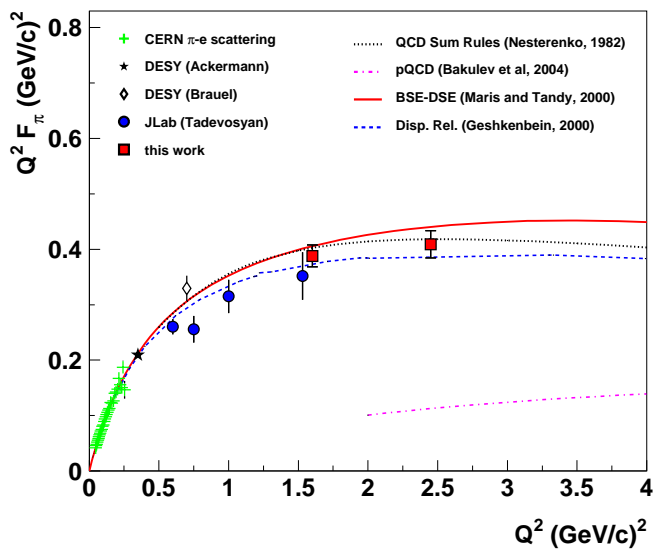}
\caption{\label{fig-fpi} \it Pion form factor as extracted in this work. Also shown are $e-\pi$ elastic data from CERN, and earlier pion electroproduction data from DESY and Jefferson Lab. The earlier Jefferson Lab data are taken from reference \cite{Tad06}. The data point at $Q^2=1.60$ GeV$^2$ from \cite{Tad06} has been shifted from its central value for visual representation. The curves are from a Dyson-Schwinger equation (solid, \cite{Mar00}), QCD sum rules (dotted, \cite{Nes82}), dispersion relations with QCD constraint (dashed, \cite{Ges00}), and from a pQCD calculation (dashed-dotted, \cite{Bak04}).}
\end{center}
\end{figure}

In Figure~\ref{fig-fpi}, our results are shown along with results from CERN, DESY, earlier Jefferson Lab data, and some representative calculations. Comparing the result at $Q^2=1.60$ GeV$^2$ to the earlier Jefferson Lab data point at a lower value of $W$ allows for a direct test of the theoretical model dependence. A higher value of $W$ allows for a measurement at smaller values of $-t$, at closer proximity to the pion pole. The data are consistent with the previous Jefferson Lab $F_{\pi}$ measurement at a value of $Q^2=1.60$ GeV$^2$ and suggest a small model uncertainty due to fitting the VGL model to the data. The data indicate a one sigma deviation from a monopole form factor that yields the measured charge radius. That form factor is up to $Q^2$=2.5 GeV$^2$ indistinguishable from the solid curve in Figure~\ref{fig-fpi}. Various models provide a good description of the measured values for $F_{\pi}$ up to $Q^2$=1.60 GeV$^2$. The data are well described by the calculation of Nesterenko and Radyushkin \cite{Nes82}, in which a QCD sum rule framework for the soft contribution to $F_{\pi}$ as well as an asymptotically dominant hard gluon exchange term is used. The dispersion relation calculation by Geshkenbein \cite{Ges00} also agrees well with the data. The data are also reasonably well described by the Dyson-Schwinger calculation by Maris and Tandy, which is based on the Bethe-Salpeter equation with dressed quark and gluon propagators. All parameters in the latter calculation are determined without the use of $F_{\pi}$ data \cite{Mar98, Mar00}. Perturbative QCD calculations of which one is shown in Figure~\ref{fig-fpi} give values of $Q^2 F_{\pi}$ around 0.10 GeV$^2$ in the region of our measurements.

In summary, we have measured separated $^{1}$H(e,e$^\prime \pi^+$)n cross sections at values of $Q^2$=1.60 and 2.45 GeV$^2$ at $W$=2.22 GeV. The charged pion form factor was extracted from the separated longitudinal cross section using a Regge model. The data are consistent with the previous Jefferson Lab result at $Q^2=1.60$ GeV$^2$. The data deviate by one sigma from a monopole form factor obeying the measured charge radius, but are still far from the values expected from pQCD calculations.

\medskip
This work was supported in part by the U.S. Department of Energy. The Southeastern Universities Research Association (SURA) operates the Thomas Jefferson National Accelerator Facility for the United States Department of Energy under contract DE-AC05-84150. We acknowledge additional research grants from the U.S. National Science Foundation, the Natural Sciences and Engineering Research Council of Canada (NSERC), NATO, and FOM (Netherlands).


\begin{thebibliography}{22}
\bibitem{Farr79} G.R.~Farrar and D.R.~Jackson, Phys. Rev. Lett. \textbf{43}, 
246 (1979).
\bibitem{Isg84} N.~Isgur and C.H.~Llewellyn-Smith, Phys. Rev. Lett. \textbf{52},
1080 (1984); N.~Isgur and C.H.~Llewellyn-Smith, Nucl. Phys. \textbf{B317},
526, (1989).
\bibitem{Ame84} S.R.~Amendolia et al., Phys. Lett. \textbf{B138}, 
454 (1984); Nucl. Phys. \textbf{B277}, 168 (1986).
\bibitem{Beb76} C.J.~Bebek et al., Phys. Rev. \textbf{D13}, 
25 (1976).
\bibitem{Beb78} C.J.~Bebek et al., Phys. Rev. \textbf{D17}, 
1693 (1978).
\bibitem{Bra77} P.~Brauel et al., Phys. Lett. \textbf{B69}, 
253 (1977); P.~Brauel et al., Z. Phys. \textbf{C3}, 
101 (1979).
\bibitem{Ack78} H.~Ackermann et al., Nucl. Phys. \textbf{B137}, 
294 (1978).
\bibitem{Vol01} J.~Volmer et al., Phys. Rev. Lett \textbf{86},
1713 (2001).
\bibitem{Tad06} V.~Tadevosyan et al., submitted to Phys. Rev. C as Rapid Communication,
 (2006).
\bibitem{Horn06} T.~Horn, Ph.D. thesis, University of Maryland (2006).
\bibitem{Asa05} R.~Asaturyan et al., Nucl. Instrum. Meth.\textbf{A548},
364-374 (2005).
\bibitem{Arr04} J.~Arrington, Phys. Rev. \textbf{C69}, 
022201 (2004).
\bibitem{Van98} M.~Vanderhaeghen, M.~Guidal and J.-M.~Laget, Phys. Rev. \textbf{C57},
1454 (1998); Nucl. Phys. \textbf{A627} 645 (1997).
\bibitem{Nes82} V.A.~Nesterenko and A.V.~Radyushkin, Phys. Lett. \textbf{B115},
410 (1982).
\bibitem{Ges00} B.V.~Geshkenbein, Phys. Rev. \textbf{D61}, 
033009 (2000).
\bibitem{Mar98} P.~Maris and C.D.~Roberts, Phys. Rev. \textbf{C58},
3659 (1998).
\bibitem{Mar00} P.~Maris and P.C.~Tandy, Phys. Rev. \textbf{C62},
204 (2000).
\bibitem{Bak04} A.P.~Bakulev et al., Phys. Rev. \textbf{D70},
033014 (2004).
\end{thebibliography}
\end{document}